\documentstyle[sprocl]{article}

\input epsf

\def\Journal#1#2#3#4{{#1} {\bf #2}, #3 (#4)}

\def\NPB{{\em Nucl. Phys.} B}
\def\PLB{{\em Phys. Lett.}  B}
\def\PRL{\em Phys. Rev. Lett.}
\def\PRD{{\em Phys. Rev.} D}

\def\be{\begin{equation}}
\def\ee{\end{equation}}
\def\beq{\begin{eqnarray}}
\def\eeq{\end{eqnarray}}

\begin{document}
\begin{flushright}
NORDITA 98/49 HE
\end{flushright}

\vspace{1cm}

\title{ BPS AND NON--BPS DOMAIN WALLS IN SUPERSYMMETRIC QCD}

\author{A.V. SMILGA}

\address{NORDITA, Blegdamsvej 17, DK-2100, Copenhagen $/\!\!\!O$, Denmark
\footnote{Permanent address: 
ITEP, B. Cheremushkinskaya 25, Moscow 117218, Russia}}


\maketitle\abstracts{We study the spectrum of the domain walls
interpolating between different chirally asymmetric vacua
in supersymmetric QCD with the $SU(N)$ gauge group and
including $N-1$ pairs of chiral matter multiplets in fundamental
and anti-fundamental representations.
There are always ``real walls'' interpolating between the chirally symmetric 
and a chirally asymmetric vacua which are BPS saturated.
 For small enough masses, there
 are two different ``complex''  BPS wall solutions interpolating between 
different chirally asymmetric vacua
 and two types of ``wallsome sphalerons''.
  At some $m = m_*$, two BPS branches join together and, in some interval
 $m_* < m < m_{**}$,
 BPS equations have no solutions, but there are solutions to the
  equations of motion
describing a non--BPS domain wall and a sphaleron. For $m > m_{**}$,
there are no complex wall solutions whatsoever.}

\section{Introduction}

Supersymmetric QCD is the theory involving  a gauge
 vector supermultiplet
$V$ and some number of chiral matter supermultiplets.
The models of this class attracted  attention of theorists since the
beginning of the
eighties and many interesting and non--trivial results concerning their
non--perturbative dynamics  have been obtained \cite{brmog}.
The dynamics depends in an essential way on the gauge group, the matter
content, the masses of the matter fields and their Yukawa couplings.

 The most simple  in some sense variant of the
model is based on the $SU(N)$ gauge group and involves $N-1$ pairs of chiral
matter supermultiplets
 $S_{i\alpha}$, $S_i^{'\alpha}$
in the fundamental and anti-fundamental representations of the gauge
group with a common mass $m$.
 The lagrangian of the model reads
\beq
{\cal L} = \left( \frac{1}{4g^2} \mbox{Tr} \int d^2\theta \ W^2\ + \
{\rm H.c.}
\right)\ + \  \sum_{i=1}^{N-1} \left[ \frac{1}{4}\int d^2\theta d^2\bar\theta\
\bar S_{i}  e^V S_{i}  \right. \nonumber \\
+ \left.  \frac{1}{4}\int d^2\theta d^2\bar\theta\
 S'_{i}  e^{-V} \bar S'_{i}
 - \frac m2 \left(  \int d^2\theta\  S'_i S_{i}
+\mbox{H.c.}\right) \right]\ ,
\label{LSQCD}
\eeq
  color and Lorentz indices are suppressed. In this case,
 the gauge symmetry is broken
completely and the theory involves a discrete set of vacuum states.
The presence of $ N$ chirally asymmetric  states
has been known for a long time. They are best seen in the weak coupling
limit $m \ll \Lambda_{SQCD}$ where the chirally asymmetric states involve
large vacuum expectation values of the squark fields $\langle s_i\rangle  \ 
\gg \Lambda_{SQCD}$
and the low energy dynamics of the model is described in terms of the colorless
composite fields ${\cal M}_{ij} = 2S'_i S_j$. The effective lagrangian
presents a Wess--Zumino model with the superpotential
 \beq
  \label{Higgs}
  {\cal W} = - \frac 23 \frac{\Lambda_{\rm SQCD}^{2N + 1}}
  {{\rm det} {\cal M}} - \frac{m}{2} {\rm Tr}\ {\cal M}
  \eeq
  The second term in Eq.(\ref{Higgs}) comes directly from the lagrangian
  (\ref{LSQCD}) and the first term is generated dynamically by instantons.
  Assuming ${\cal M}_{ij} = X^2 \delta_{ij}$ and solving the equation
  $\partial {\cal W}/\partial \chi = 0$ ($\chi$ is the scalar component of the
  superfield $X$), we find $N$ asymmetric vacua
  \beq
  \label{vacchi}
  \langle\chi\rangle_k = \left( \frac 43 \frac{\Lambda_{\rm SQCD}^{2N+1}} m
  \right)^{1/2N}
e^{\pi i k/N} \ \equiv \ \ \rho_* e^{\pi i k/N}
  \eeq
  (the vacua ``k'' and ``k + N''  have the same value of the
  moduli $\langle\chi^2\rangle_k$ and  are physically equivalent).
  These vacua are characterized by a finite gluino condensate
  \beq
  \label{cond}
  \langle{\rm Tr}\ \lambda^2\rangle_k \  = \ 8\pi^2  m\langle\chi^2\rangle_k
  \eeq

 It was noted recently \cite{Kovner} that on top of (\ref{vacchi})
  also a chirally symmetric vacuum with the
zero value of the condensate exists.  It cannot be detected
in the framework
of Eq.(\ref{Higgs}) which was derived {\it assuming} that the scalar v.e.v.
and the gluino condensate are nonzero and large, but is clearly
seen if writing down
the effective lagrangian due to Taylor, Veneziano, and
Yankielowicz (TVY) \cite{TVY} involving also the composite field
 \beq
 \label{Phi}
\Phi^3 = \frac 3{32\pi^2} {\rm Tr}\ W^2
 \eeq
   The corresponding superpotential reads

  \beq
  \label{TVY}
{\cal W} =  \frac 23 \Phi^3 \left[ \ln \frac{\Phi^3 X^{2(N-1)}}
{\Lambda_{SQCD}^{2N + 1}} \ -\ 1 \right] - \frac{m}{2} N X^2
\eeq
  
Choosing kinetic term in the simplest possible form
 \beq
\label{supkin}
{\cal L}_{\rm kin} \ =\ \int d^4\theta (\bar X X + \bar \Phi \Phi) ,
 \eeq
the corresponding scalar potential   is
 \beq
U(\phi, \chi) \ =\ \left|\frac{\partial {\cal W}}{\partial \phi}
\right|^2 +
 \left|\frac{\partial {\cal W}}{\partial \chi}\right|^2 \ =\
4\left| \phi^2 \ln\{\phi^3 \chi^{2(N-1)}\} \right|^2 + (N-1)^2
\left|m\chi  - \frac{4\phi^3}{3\chi} \right|^2
\label{potTVY}
  \eeq
(from now on we set $\Lambda_{\rm SQCD} \equiv 1$).
  The potential (\ref{potTVY}) has $N+1$ degenerate minima. One of them is
  chirally
symmetric: $\phi = \chi = 0$.  There are also $N$ chirally asymmetric vacua
with $\langle\chi\rangle_k$ given in Eq.(\ref{vacchi}) and
   \beq
\langle\phi\rangle_k \ =\ \left( \frac {3m}4 \right)^{(N-1)/(3N)}
  e^{-\frac{2i(N-1)\pi k}{3N}}\ \equiv \ R_*  e^{-\frac{2i(N-1)\pi k}{3N}}
 \label{vacphi}
  \eeq

The presence of different degenerate physical vacua in the theory
implies the existence of domain walls --- static field configurations
depending  only on one spatial coordinate ($z$) which interpolate between
one of the vacua at $ z = -\infty$ and another one at $z = \infty$ and
minimizing the energy functional. As was shown in \cite{Dvali}, in many
cases the energy density of these walls can be found exactly due to the
fact that the walls present the BPS--saturated states.

The energy density of a BPS--saturated wall in SQCD with $SU(N)$ gauge
group satisfies a relation \cite{my}
  \beq
   \label{eps}
   \epsilon \ =\ \frac {N}{8\pi^2} \left|\langle{\rm Tr}\ \lambda^2\rangle_\infty
    \ -\ \langle{\rm Tr}\ \lambda^2\rangle_{-\infty} \right|
    \eeq
where the subscript $\pm \infty$ marks the values of the gluino
condensate at spatial infinities.

 Bearing Eqs. (\ref{eps},\ \ref{cond}) in mind , the energy densities of the
BPS walls are
  \beq
 \label{epsr}
 \epsilon_r \ =\ N \left( \frac {4m^{N-1}}3  \right)^{1/N}
  \eeq
  for the  real walls interpolating between a chirally asymmetric
and the chirally symmetric vacua (we call them ``real'' because
the corresponding gauge field configurations are essentially real) and
  \beq
  \label{epsc}
\epsilon_c = 2 \epsilon_r  \sin \frac \pi N
 \eeq
 for the complex walls interpolating between different chirally asymmetric
vacua. 
 The RHS of Eqs.(\ref{eps}-\ref{epsc}) presents an
absolute   lower bound for the energy of {\it any} field configuration
interpolating between different vacua.

    The relation (\ref{eps}) is valid {\it assuming} that the wall exists and
is BPS--saturated. However, whether such a BPS--saturated domain wall
exists or not is a non--trivial dynamic question which can be answered
only in a specific study of a particular theory in interest. This talk 
presents a brief review of our papers
\cite{my}--\cite{SV} where such a study has been performed. 

 The results concerning the BPS walls were obtained by solving numerically
 the first order BPS equations
   \beq
  \partial_z \phi \ =\ e^{i\delta} \partial \bar {\cal W} /\partial \bar
\phi,  \ \ \ \ \      \partial_z \chi \ =\ e^{i\delta}
\partial \bar {\cal W} /\partial \bar \chi\
  \label{BPS}
  \eeq
associated with the TVY lagrangian. 
The value of $\delta$ to be chosen in Eq.(\ref{BPS}) depends on the wall 
solution we are going to find. To fix it, note that the equations (\ref{BPS})
admit an integral of motion:
\beq
\label{ImW} 
{\rm Im} [ {\cal W}(\phi, \chi)e^{-i\delta}] \ =\ {\rm const}
\eeq
Indeed, we have
$$
e^{-i\delta} \partial_z {\cal W} = e^{-i\delta} \left(
\frac{\partial {\cal W}}{\partial \phi} \partial_z \phi  \ + \ 
\frac{\partial {\cal W}}{\partial \chi} \partial_z \chi \right) \ =  \\ 
\left| \frac{\partial {\cal W}}{\partial \phi} \right |^2 \ +\ 
\left| \frac{\partial {\cal W}}{\partial \chi} \right|^2 \ =\ 
e^{i\delta} \partial_z \bar {\cal W}
$$
  The real wall connects the vacua with ${\cal W} = 0$ and ${\cal W} = 
-N/2 (4m^{N-1}/3)^{1/N}$. These boundary conditions are consistent with 
Eq.(\ref{ImW}) only if $e^{i\delta} = \pm 1$ (the sign depends on whether the
walls goes from the symmetric vacuum to the asymmetric one when $z$ goes
from $-\infty$ to $+\infty$ or the other way round). For the complex walls,
the boundary conditions and the conservation law (\ref{ImW}) imply $\delta = 
\frac {\pi}N \pm \frac {\pi}2$.

To study the spectrum of the domain walls which are not BPS--saturated, one
has to solve the equations of motion which are of the second order, and,
technically, the problem is a little bit more involved.

\section{Real Walls}
The BPS equations (\ref{BPS}) with the superpotential (\ref{TVY})
and the kinetic term (\ref{supkin}) have the form
  \beq
\label{fihi}
 \phi '\  = \ e^{i\delta} \cdot 2\bar \phi^2 \ln\{\bar\phi^3
\bar \chi^{2(N-1)}
 \} \nonumber \\
 \chi ' \ =\ e^{i\delta} \cdot (N-1) \left[
 \frac{4\bar\phi^3}{3\bar\chi} - m\bar\chi \right]
  \eeq
($O' \equiv \partial_z O$).
To find the wall interpolating between $\phi = \chi = 0$ at $z = -\infty$
and $\phi = R_*, \chi = \rho_*$ at $z = \infty$, we have to choose
$\delta = \pi$ (or $\delta = 0$ for the wall going in the opposite direction).
With this choice and the boundary conditions given, the solutions $\phi(z)$
and $\chi(z)$  are going to be real so that we have a simple system of just
two first--order differential equations.

The dynamics of this system is essentially the same for any $N$.
 The solution exists for all
masses. For large $m$, the heavy matter field can be integrated out, and we
arrive at the BPS equation for the supersymmetric gluodynamics
  \beq
 \label{fiN}
\phi' \ =\ -2N \phi^2 \ln \left(\phi^3/R_*^3\right)
  \eeq
The solution of this equation with the boundary conditions
 $\phi(-\infty) = 0,\ \phi(\infty) = R_*$ can be expressed into integral
logarithms.

\begin{figure}
{
    \begin{center}
        \epsfxsize=300pt
        \epsfysize=200pt
        \vspace{-17mm}
        \parbox{\epsfxsize}{\epsffile{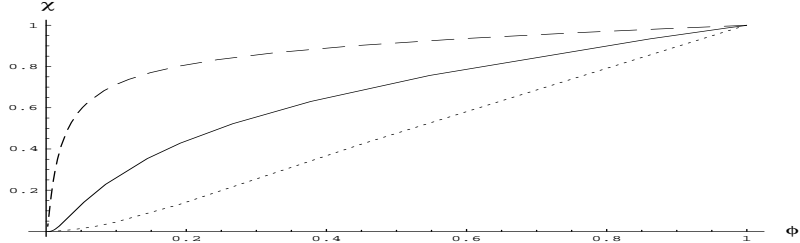}}
        \vspace{-15mm}
    \end{center}
}
\caption{ Real walls ($N=3$). The parametric plots for $m=1$ (solid line),
$m = .1$ (dashed line), and $m = 10$ (dotted line). $\phi$ and $\chi$ are
measured in units of $R_*$ and $\rho_*$, respectively. \label{reals}}
\end{figure}

An analytic solution of the system (\ref{fihi}) can be
found  also for small masses. 
In the intermediate range of masses, the solution has to be found numerically.
Parametric plots in the ($\phi, \ \chi$) plane for $N = 3$ and 
 different values of $m$ are drawn in Fig. \ref{reals}.

 \section{Domain Walls in Higgs Phase}
\begin{figure}
{
    \begin{center}
        \epsfxsize=300pt
        \epsfysize=200pt
        \vspace{-17mm}
        \parbox{\epsfxsize}{\epsffile{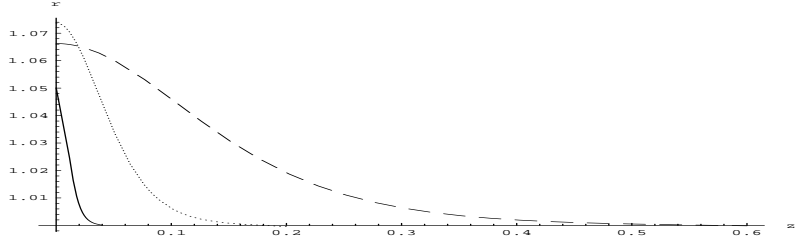}}
        \vspace{-15mm}
    \end{center}
}
\caption{ BPS walls in Higgs phase for $N = 3$ (dashed line), $N = 5$
(dotted line) and $N = 10$ (solid line).  \label{prof0s}}
\end{figure}

The lagrangian (\ref{Higgs}) can be obtained from the full TVY lagrangian
(\ref{TVY}) in the limit of small masses if integrating out the heavy gauge 
{\it under the assumption}  that the Higgs expectation value of
the scalar matter fields is large. In this approach, chirally symmetric vacuum 
is not seen and there are only complex walls which turn out to be BPS 
saturated (we have seen, however, that the chirally symmetric vacuum and the 
real walls
associated with it exist actually for all masses no matter how small they are).
To find the  profile of complex walls in the Higgs phase, we have got to 
solve the BPS equations
  \beq
  \label{BPSHig}
 \chi ' \ =\ e^{-i \pi(N-2)/2N} 
\frac{\partial \bar {\cal W}}{\partial \bar \chi}\, ,
    \eeq
where ${\cal W}$ is the superpotential (\ref{Higgs}), with the boundary
conditions
 \beq
 \label{bcHgs}
\chi(-\infty) \ =\ \rho_*;
\ \ \ \ \ \  \chi(\infty) \ =\  \rho_*e^{i\pi/N}\, ;
  \eeq
  It is convenient to introduce polar variables $\chi \ =\ \rho e^{i\alpha}$. 
The equations (\ref{BPSHig}) acquire the form
  \beq
\label{rogam}
 \rho ' \ =\ (N-1) \left\{ m\rho \sin \gamma \ -\ \frac 4{3\rho^{2N-1}}
\sin [\gamma(N-1)] \right\} \nonumber \\
 \gamma '\ =\ 2(N-1) \left\{ m \cos \gamma \ + \ \frac 4{3\rho^{2N}}
\cos [\gamma(N-1)] \right\}
  \eeq
 where $\gamma \equiv 2\alpha - \pi/N$ changes from $\gamma = -\pi/N$ at
$z = -\infty$ to $\gamma = \pi/N$ at $z = \infty$.
 For $N = 2$, the solution is
analytic:
  \beq
  \label{rogamsol}
\rho(z) \ &=&\ \rho_* \nonumber \\
\tan \gamma(z) \ &=&\ \sinh [4m(z-z_0)]
  \eeq
or in the complex form:
  \beq
 \label{chisol}
\chi(z) \ =\ \rho_* \frac {1 + i e^{4m(z - z_0)}}{\sqrt{1 + e^{8m(z-z_0)}}}
  \eeq
($z_0$ is the position of the wall center).
For $N \geq 3$, the solutions can be found numerically. The profiles for the
ratio
$r(z) = \rho(z)/\rho_*$ in the interval $z_0 \equiv 0 \leq z < \infty$
[it is a half of the wall, another half being restored by symmetry
considerations: $\rho(-z) = \rho(z)$]
with different $N = 3, 5 , 10$ are presented in Fig. \ref{prof0s}.

We see that the dependence $\rho(z)$ is not flat anymore but displays a bump
in the middle. To understand it, remind that the system (\ref{rogam})
has the integral of motion (\ref{ImW}). In our case, it amounts to
  \beq
 \label{intmot}
\frac{m(N-1)}2 \rho^2 \cos \gamma  - 
\frac2{3 \rho^{2(N-1)}} \cos [ \gamma (N-1)]   =  
\frac N2 \left( \frac{4m^{N-1}}3 \right)^{1/N} \cos \frac \pi N 
   \eeq
as follows from the boundary conditions (\ref{bcHgs}) with 
$\delta = \pi/N - \pi/2$. In the middle of the wall, $\gamma = 0$, 
and  the condition
(\ref{intmot}) implies
 \beq
  \label{xN}
(N-1) x^2 - \frac 1{x^{2(N-1)}} \ =\ N \cos \frac \pi N
  \eeq
( $x \equiv r(0)$). It is not difficult to
observe that the real root of the algebraic equation (\ref{xN}) is slightly
greater than 1 for $N \geq 3$. When $N$ is large, $x-1$ tends to zero 
$\propto 1/N$.

 \section{Two BPS solutions and Phase Transition in Mass}

When $m \neq 0$, the gauge degrees of freedom associated with the superfield
$\Phi$ do not decouple completely and should be taken into account. The full
system (\ref{fihi}) of the BPS equations for the complex domain walls has the
form
   \beq
   \label{4sys}
    \begin{array}{l}
 \rho '\ =\ (N-1) \left[ m\rho \sin (2\alpha - \pi/N) -
 \frac{4R^3}{3\rho} \sin (3\beta - \pi/N) \right]
\\
 \alpha '\ =\ (N-1) \left[ m \cos (2\alpha - \pi/N) -  
\frac{4R^3}{3\rho^2} \cos (3\beta - \pi/N) \right]
 \\
 R '\ =\ -2R^2 \left[ \sin (3\beta - \pi/N) \ln(R^3\rho^{2(N-1)}) +
\cos (3\beta - \pi/N) [3\beta + 2\alpha (N-1)] \right]  \\
 \beta '\ =\ 2R \left[- \cos(3\beta - \pi/N) \ln(R^3\rho^{2(N-1)}) +
\sin (3\beta - \pi/N) [3\beta + 2\alpha (N-1)] \right] 
\end{array} 
   \eeq
 where, as in the previous section, we have chosen $\delta \ =\ \pi/N -\pi/2$
and introduced the polar variables $\chi = \rho e^{i\alpha},\ 
\phi = R e^{i\beta}$. One should solve the system (\ref{4sys}) with the 
 boundary conditions
  \beq
  \label{4bc}
\rho(-\infty)  =\rho(\infty) = \rho_*; \ \ R(-\infty) = R(\infty) = R_*; 
\nonumber \\
\alpha(-\infty) = \beta(-\infty) = 0;\ \ \alpha(\infty) = \pi/N;\ \ 
\beta(\infty) = - \frac{2(N-1) \pi}{3N}
  \eeq

In the limit $m \to 0$, we can just freeze the heavy variables: 
   \beq
 \label{freeze}
\beta = 
-2\alpha(N-1)/3;\ R = \rho^{-2(N-1)/3}
  \eeq
 in which case the first two equations 
in Eq.(\ref{4sys}) reproduce the system (\ref{rogam}) studied in the previous 
section. When mass is nonzero, but small enough, it is possible to develop
a systematic  Born--Oppenheimer expansion and to find the profile of the wall
as a series over the small parameter $m$ (or rather $m^{(2N+1)/3N}$ in our 
case).
Not going in details \cite{SUN}, we just present here for illlustration
how $R(0)$, the absolute value of the field $\phi$ in the center of the wall,
is changed with mass:
  \beq
 \label{etam}
\eta(m) \ =\ \frac{R(0)}{R_0(0)} \ =\ 1\  - \ \frac{(N-1)^2}{9R_0(0)} 
\left[ m + \frac 4{3\rho_0^{2N}(0)} \right] - \nonumber \\
\frac{(N-1)^3}{162 R_0^2(0)} 
\left[ m + \frac 4{3\rho_0^{2N}(0)} \right]
\left[ m(7N-1) - \frac{32(N-1)}{3\rho_0^{2N}(0)} \right] \ + \nonumber \\
O[m^3/R_0^3(0)]
  \eeq
where $R_0(0) \propto m^{(N-1)/(2N+1)}$ is related to 
$\rho_0(0) = x\rho_*$ as in 
Eq.(\ref{freeze}), and $\rho_0(0)$ was found earlier when we studied the 
solutions of the reduced system in the Higgs phase.
 
\begin{figure}
{
    \begin{center}
        \epsfxsize=300pt
        \epsfysize=150pt
        \vspace{2mm}
        \parbox{\epsfxsize}{\epsffile{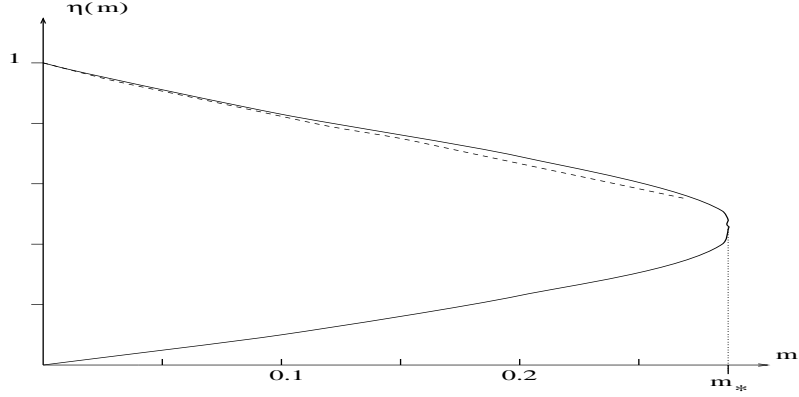}}
        \vspace{1mm}
    \end{center}
}
\caption{ The 
ratio $\eta = R(0)/R_0(0)$  as a function of mass for the 
$SU(3)$ theory. The dashed
line describes the analytic result (\ref{etam}) valid for small masses.
  \label{plot3s}}
\end{figure}

  When $m$ is neither too large nor too small, the Born--Oppenheimer
approximation does not apply and we are in a position to solve the full 
system of 4 equations (\ref{4sys}) numerically. We did it for $N = 2,3,4$.
The results turned out to be rather nontrivial and surprising. Namely, there
turned out to be not one but {\it two} different BPS solutions. But they exist
only for small enough masses; if the mass exceeds some critical value $m_*$,
the BPS equation system has no solution at all. A kind of phase transition in
mass occurs ! The values of the critical mass $m_*$ have been found 
numerically:
  \beq
\label{mcrit}
m_*^{SU(2)} = 4.67059\ldots ,  \ \ \ \ \  
m_*^{SU(3)} = .28604\ldots , \ \ \ \ \  
m_*^{SU(4)} = 0.07539\ldots .
 \eeq
 In Fig. \ref{plot3s}, we plotted the dependence of 
$\eta = R(0)/R_0(0)$ on $m$  for both branches . 
 We see that, at $m = m_*$, two branches are joined together. This {\it is}
 the reason why no solution exists at larger masses.

 The upper BPS branch approaches the Higgs phase solution for the small masses.
Speaking of the low branch, it approaches two widely 
separated real BPS walls for $N=2$ passing through the chirally symmetric 
vacuum in the middle, but for $N \geq 3$ it presents a new 
nontrivial solution with the field values in the middle of the wall passing
at some finite distance from the chirally symmetric vacuum even in the limit
$m \to 0$ \cite{SUN}.

As is seen from Eq. (\ref{mcrit}), the value of $m_*$ expressed in the units of
$\Lambda_{\rm SQCD}$ drops rapidly with $N$.
  One can understand it looking at the Born--Oppenheimer result (\ref{etam}) 
 Indeed,
the critical mass $m_*$ and the value of the mass where the Born--Oppenheimer
expansion in Eq.(\ref{etam}) breaks down should be of the same order.
For large $N$, the expansion parameter is $\kappa \sim N^2(m/R_*) \sim
N^2 m^{2/3}$. Assuming $\kappa \sim 1$, we arrive at the conclusion that
$m_*$ falls down as $$m_*(N) \ \propto\ N^{-3}$$ in the limit $N \to \infty$.

\section{Solving equations of motion}

 Every solution of the BPS equations (\ref{BPS}) is also a solution to the 
equations of motion, but not the other way round. It turns out that nontrivial
 non--BPS {\it complex} wall solutions exist. 
 To find them, we have to solve the equations of motion 
for the potential (\ref{potTVY}) with the kinetic term 
$ |\partial \phi|^2 + |\partial \chi|^2 $.
As earlier, it is convenient to introduce the polar variables 
$\chi = \rho e^{i\alpha},\
\phi = R e^{i\beta}$ after which the equations of motion
acquire the form
 \beq
 \label{eqmot}
R'' - R \beta'^2 \ &=&\ 8R^3 [L(L + 3/2) + \beta_+^2] +
(N-1)^2 \left[\frac{16R^5}{3\rho^2} - 4mR^2 \cos (\beta_-) \right]
\nonumber \\
R\beta'' + 2R'\beta'\  &=& \ 12R^3\beta_+ + 4(N-1)^2 mR^2 \sin(\beta_-)
\nonumber \\
\rho'' - \rho \alpha'^2 \ &=&\ (N-1)\frac{8R^4}\rho L + (N-1)^2
\left(m^2\rho - \frac{16R^6}{9\rho^3} \right)  \nonumber \\
\rho \alpha'' + 2\rho' \alpha' \ &=&\ (N-1) \frac{8R^4}{\rho} \beta_+ -
(N-1)^2 \frac{8mR^3}{3\rho} \sin (\beta_-) \ ,
  \eeq
where $L = \ln[R^3 \rho^{2(N-1)}], \ \beta_+ = 3\beta + 2(N-1)\alpha,
\ \beta_- = 3\beta - 2\alpha$. The  boundary conditions (\ref{4bc}) are 
assumed.

The phase space of the system (\ref{eqmot}) is 8--dimensional
and the boundary problem (which we solved by shooting method)
is somewhat more tricky than for the BPS case. In particular, the trajectories
become more and more unstable as $N$ grows. Finding the trajectory going 
from one vacuum to another is something like
lancing the ball to roll along a razorblade mountain ridge from
one top to another.
Our computer managed to do it for $N=2$ and $N=3$.

\begin{figure}
{
    \begin{center}
        \epsfxsize=300pt
        \epsfysize=150pt
        \vspace{-1mm}
        \parbox{\epsfxsize}{\epsffile{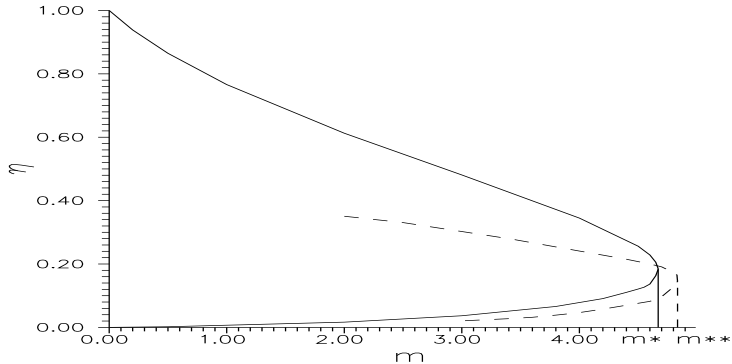}}
        \vspace{-4mm}
    \end{center}
}
\caption{ The
ratio $\eta = R(0)/R_0(0)$  for the solutions of the equations of motion
as a function of mass for the
$SU(2)$ theory. The solid
lines describe the BPS solutions and the dashed lines describe the
non--BPS wall and the sphalerons. \label{vse2}}
\end{figure}

\begin{figure}
{
    \begin{center}
        \epsfxsize=300pt
        \epsfysize=150pt
        \vspace{-1mm}
        \parbox{\epsfxsize}{\epsffile{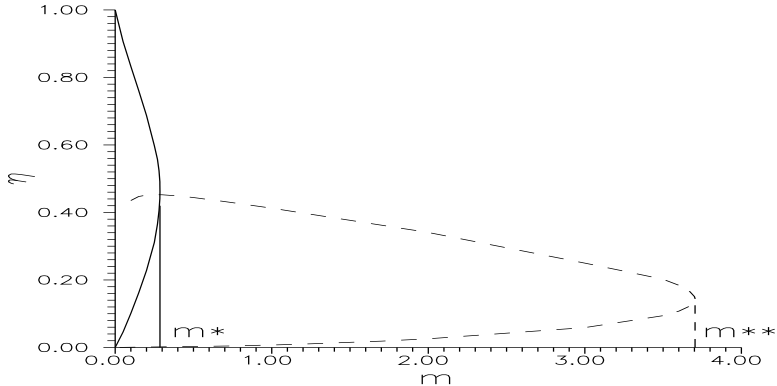}}
        \vspace{-3mm}
    \end{center}
}
\caption{ The same  for $N = 3$.  \label{vse3}}
\end{figure}

All the solutions obtained  are presented in Fig. \ref{vse2} ($N=2$) and in 
Fig. \ref{vse3} ($N=3$) where the parameter $R(0)$
is plotted as a function of mass [we normalized $R(0)$ at its value
$R_0(0) = R_*$ ($N=2$) and 
$R_0(0) = .918\ldots R_* $ ($N=3$) for the upper BPS branch in the limit
$m \to 0$ ].  For small masses, there are several
solutions. We obtain first of all the BPS solutions
studied before
(solid lines in Figs. \ref{vse2}, \ref{vse3}). We find also two new
solution branches drawn with the dashed lines. We see that,
similarly to the BPS branches, two new dashed branches fuse together at some
value $m = m_{**} > m_*$ . No solution for the system (\ref{eqmot}) exist at
$m > m_{**}$.

\begin{figure}
{
    \begin{center}
        \epsfxsize=300pt
        \epsfysize=150pt
        \vspace{1mm}
        \parbox{\epsfxsize}{\epsffile{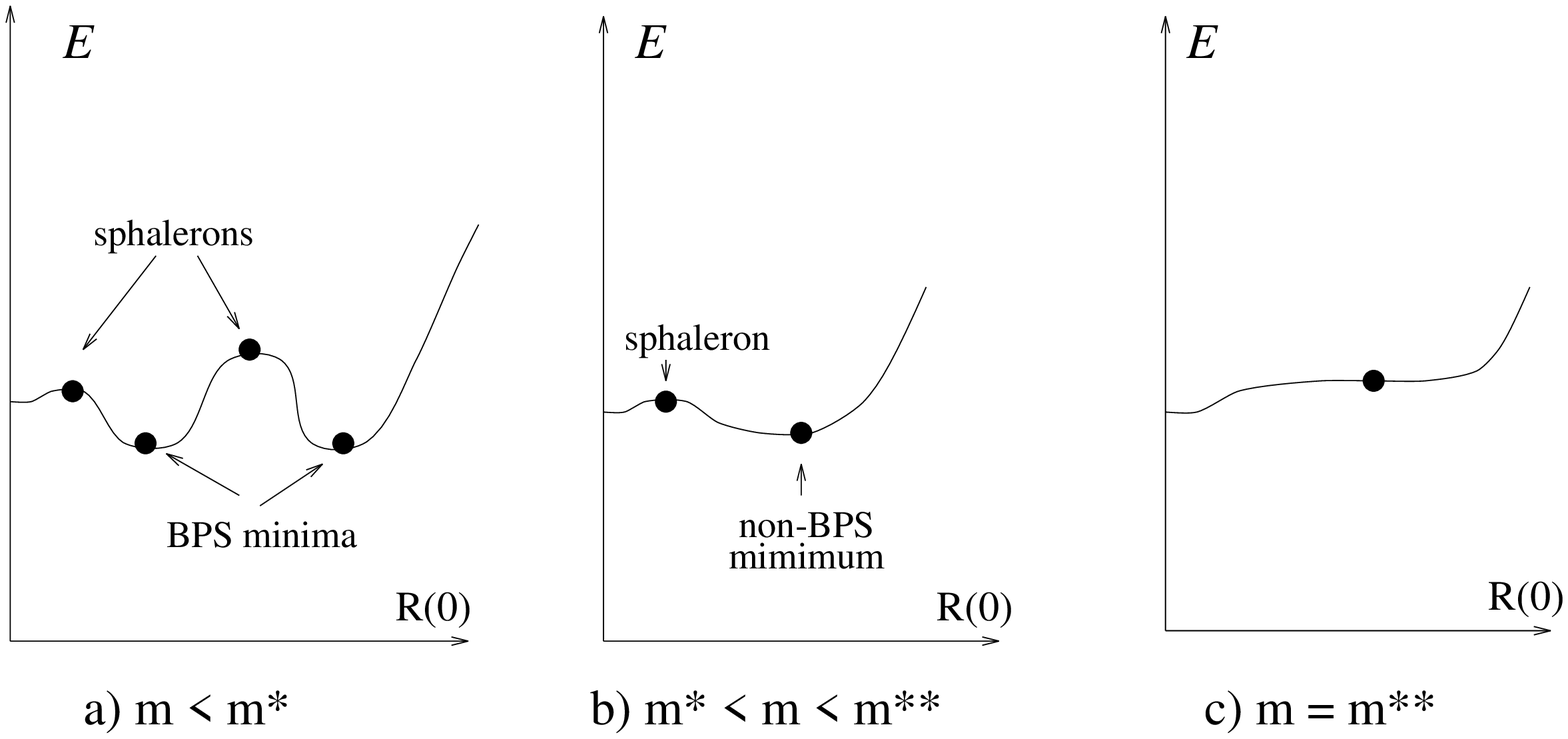}}
        \vspace{0mm}
    \end{center}
}
\caption{ Illustrative profiles of the energy functional vs. $R(0)$.
 \label{Eplot}}
\end{figure}

We see that the pattern of the solutions for $N=2$ and $N = 3$ 
is qualitatively similar.
To understand it, assume first that
$m < m_*$ and draw the energy functional $ E$ for field configurations
with wall boundary conditions minimized over all parameters except the value
 of $R(0)$  which is kept fixed (see Fig. \ref{Eplot}a ). For very small
$R(0)$, our configuration
nearly passes the chirally symmetric minimum and the minimum
of the energy corresponds to two widely separated real walls. Thus
$ E[R(0) = 0] = 2\epsilon_r$ with $\epsilon_r$ given in Eq.(\ref{epsr}).
Two minima in Fig. \ref{Eplot}a correspond to BPS solutions with the
energy $\epsilon_c$. They are separated
by an energy barrier (for illustrative 
purposes, the hight of the barriers is exagerated). The top of this barrier 
(actually, this is a saddle
point with only one unstable mode corresponding to $R(0)$, in other words
--- a {\it sphaleron})
is a solution described by the upper dashed line in Figs. \ref{vse2}, 
\ref{vse3}. The lower
dashed line corresponds to the local maximum on the energy barrier separating
the lower BPS branch and the configuration of two distant real walls at
$R(0) = 0$.
\footnote{For $N=3$, in contrast to the case $N=2$, the existence of such a 
barrier
could not be established from the BPS spectrum alone. The matter is that,
while $2\epsilon_r = \epsilon_c$ for $N=2$ and the presence of the maximum
is guaranteed by the Roll theorem, $2\epsilon_r > \epsilon_c$  for $N = 3$,
 and one could in principle imagine a situation where
$ E$ falls down monotonically when $R(0)$ is increased from zero
up to its value at the lower BPS branch. Our
numerical results strongly suggest, however, that the energy barrier (
though a tiny one) is present. }

 At $m = m_*$,
two BPS minima fuse together and the energy barrier separating them
disappears. The upper sphaleron branch coincides
with the BPS solution at this point. When $m$ is increased above $m_*$, the
former BPS minimum  is still a  minimum of the energy functional, but its
 energy is now above the BPS bound (see Fig.\ref{Eplot}b).
The corresponding solution is described by the analytic continuation of the
upper sphaleron branch. The lower dashed branch in the region
$m_* < m < m_{**}$ is still a sphaleron. At the second critical point
$m = m_{**}$, the picture is changed again (see Fig.\ref{Eplot}c). The local
maximum and the local minimum fuse together and
the only one remaining stationary point does not correspond to an extremum
of the energy functional anymore. At larger masses, no non-trivial stationary
 points are left.

There is one distinction, however. For $N=2$, the second critical mass
$m_{**} \approx 4.83$ is very close to $m_*$ and the energy of the non--BPS
wall at $m = m_{**}$ deviates from the BPS limit by less than $10^{-5}$ !
When $N=3$, $m_{**} \approx 3.704$ differs essentially from $m_*$ and the 
energy of the wall at $m = m_{**}$, being very close to the energy of two 
widely separated real walls $2\epsilon_r$, differs from the BPS limit 
$\epsilon_c$ for
the complex walls. One can speculate that, while $m_*$ drops rapidly as $N$
grows, the value of $m_{**}$ is roughly constant in the large $N$ limit.
To be precise,  we did not perform a careful numerical
study for $N > 3$ and cannot therefore exclude the possibility that, 
for large 
enough $N$, the second critical mass disappears altogether and the non-BPS 
complex wall solutions exist for any mass. Our bet, however, is that it does 
not happen.

\section{Discussion.}

Our main result is that, while the real BPS domain walls connecting 
the chirally symmetric and a chirally asymmetric vacua are present at
all masses, the complex BPS walls interpolating between different asymmetric
vacua exist only for small enough masses $m < m_*$, $m_*$ being
 given in Eq.(\ref{mcrit}).
  A kind of phase transition associated with the restructuring of the wall
spectrum occurs.
\footnote{Needless to say, it is not  a phase transition of a habitual
 thermodynamic variety. In particular, the vacuum energy is zero both
  below and above the phase transition point --- supersymmetry is never
  broken here. Hence $E_{vac} (m) \equiv 0$ is not singular at 
$m = m_*$.}

Furthermore, 
 also a second critical mass $m_{**}$ exists beyond
which no complex wall solution can be found whatsoever. 
That means in particular that no domain walls connecting different chiraly
asymmetric vacua are left in the pure supersymmetric Yang--Mills theory
corresponding to the limit $m \to \infty$, and only the real domain walls
connecting the chirally symmetric and a chirally asymmetric vacuum states
survive in this limit. Note that that contradicts a recent {\it assumption} 
by Witten \cite{Witten} that it is complex rather than real
domain walls which are present in the pure SYM theory (Witten discussed them
in the context of brane dynamics).

One has to
make a reservation here: our result was obtained in the framework of the
TVY effective lagrangian (\ref{TVY}) whose status [in contrast to that
 of the lagrangian (\ref{Higgs})] is not absolutely clear:
the field $\Phi$ describes heavy  degrees of freedom (viz. a scalar
glueball and its superparnter) which are not nicely separated
from all the rest. However, the form of the superpotential (\ref{TVY})
and hence the form of the lagrangian for static field configurations
is {\it rigidly} dictated by symmetry considerations; the uncertainty
involves only kinetic terms. It is reasonable to assume that, as far as
the vacuum structure of the theory is concerned (but not e.g. the excitation
spectrum --- see Ref.\cite{SV} for detailed discussion), the effective TVY
potential (\ref{TVY}) can be trusted. A recent argument against using
Eq. (\ref{TVY}) that the chirally symmetric phase whose existence follows
from the TVY lagrangian does not fulfill certain discrete anomaly matching
conditions \cite{Csaki} is probably not sufficient. First, it assumes that
the excitation
spectrum in the symmetric phase is the same as it appears in the TVY
lagrangian which is not justified. Second, it was argued  recently
that the TVY lagrangian describes actually {\it all} the relevant symmetries
of the underlying theory and the absence of the anomaly matching is in a sense
an ``optical illusion'' \cite{KKS}.

It would be very interesting to repeat the numerical study performed here
for the effective lagrangian with the same TVY superpotential (\ref{TVY}), but
a different kinetic term. 
Probably, it makes sense to start with the VY lagrangian for the pure SYM 
theory and try
 tentatively for example
 \beq
 \label{kin2}
{\cal L}_{\rm kin}' \ =\ \int d^4\theta \frac {\bar \Phi D^2 \Phi}{(\bar \Phi 
\Phi)^{1/2}}
 \eeq
(One {\it can}not write instead a general K\"ahler term 
$\int d^4\theta F(\bar \Phi, \Phi)$.
It would involve a dimensionful scale and would spoil the anomaly matching
condition $\partial {\cal L}_{\rm eff}/\partial (\ln \Lambda) \sim G^2$.).
Our {\it guess} is that, for the VY lagrangian with the kinetic term 
(\ref{kin2}), the real walls talking to the chirally symmetric vacuum would 
also be present and the complex walls would be also absent.
For a generalized TVY lagrangian, the particular values of $m_*$ and $m_{**}$ 
would be different, but 
the qualitative pictures in Figs. \ref{vse2}, \ref{vse3} would not change.

\section*{Acknowledgments}
This work was supported in part  by the RFBR--INTAS grants 93--0283, 94--2851,
95--0681, and 96-370, by the RFFI grants 96--02--17230,  97--02--17491, and
97--02--16131, by the RFBR--DRF grant 96--02--00088,
by the U.S. Civilian Research and Development Foundation under award
\# RP2--132, and by the Schweizerishcher National
Fonds grant \# 7SUPJ048716.


\section*{References}
\begin{thebibliography}{99}


\bibitem{brmog}
V. Novikov, M. Shifman, A. Vainshtein, and V. Zakharov, \Journal{\NPB}
 {229}{407} {1983}; \Journal{\PLB} {166}{334}{1986};
I. Affleck, M. Dine, and N. Seiberg, \Journal{\NPB} {241}{493} {1984} 493; 
\Journal{\NPB}{256}{557} {1985};
G. Rossi and G. Veneziano, \Journal{\PLB}
 {138}{195}{1984};
D. Amati, K. Konishi, Y. Meurice, G. Rossi, and G. Veneziano,
{\it Phys. Rep.} {\bf 162},  557 (1988);
M. Shifman,
{\it Int. J. Mod. Phys.} A {\bf 11}, 5761  (1996).

\bibitem{Kovner}
A. Kovner and M. Shifman, \Journal{\PRD} {56}{3296} {1997}.

\bibitem{TVY}
T. Taylor, G. Veneziano, and S. Yankielowicz, \Journal{\NPB}
 {218}{493} {1983}.

\bibitem{Dvali}
G. Dvali and M. Shifman,  \Journal{\PLB}{396}{64}{1997};
B. Chibisov and M. Shifman, \Journal{\PRD}  {56}{7990}{1997}.

\bibitem{my} A. Kovner, M. Shifman, and A. Smilga, \Journal{\PRD}
{56}{7978}{1997}.

\bibitem{my1} A. Smilga and A. Veselov, \Journal{\PRL}
 {79} {4529} {1997}; \Journal{\PLB} {428} {303} {1998}.


\bibitem{SUN} A. Smilga, hep-th/9711032 , {\it Phys. Rev.} D, to appear.

\bibitem{SV} A. Smilga and A. Veselov, \Journal{\NPB}
 {515}{163}{1998}.

\bibitem{Witten} E. Witten, \Journal{\NPB} {507}{658} {1997}.

\bibitem{Csaki} C. Csaki and H. Murayama, \Journal{\NPB} {515} {114} {1998}.


\bibitem{KKS} I. Kogan, A. Kovner, and M. Shifman, 
\Journal{\PRD} {57} {5195} {1998}.

\end{thebibliography}
\end{document}